\documentclass[aps,preprint,groupedaddress,amsmath,amssymb,longbibliography]{revtex4-1}
\usepackage{amsmath}
\usepackage{graphicx}
\usepackage{dcolumn}
\usepackage{bm}
\usepackage{xcolor}

\begin{document}

\title{Long-lasting Quantum Memories: Extending the Coherence Time of Superconducting Artificial Atoms in the Ultrastrong-Coupling Regime}

\author{Roberto Stassi$^{1}$}\email{roberto.stassi@riken.jp} 
\author{and Franco Nori$^{1,2}$}

\affiliation{$^1$CEMS, RIKEN, Saitama 351-0198, Japan} 
\affiliation{$^2$Physics Department, The University of Michigan, Ann Arbor, Michigan 48109-1040, USA}

\date{\today}

\begin{abstract}
Quantum systems are affected by interactions with their environments, causing decoherence through two processes: pure dephasing and energy relaxation.  For quantum information processing it is important to increase the coherence time of Josephson qubits and other artificial two-level atoms. We show theoretically that if the coupling between these qubits and a cavity field is longitudinal and in the ultrastrong-coupling regime, the system is strongly protected against relaxation. Vice versa, if the coupling is transverse and in the ultrastrong-coupling regime, the system is protected against pure dephasing. Taking advantage of the relaxation suppression, we show that it is possible to enhance their coherence time and use these qubits as quantum memories. Indeed, to preserve the coherence from pure dephasing, we prove that it is possible to apply dynamical decoupling.  We also use an auxiliary atomic level to store and retrieve quantum information.

\end{abstract}

\maketitle

\section{Introduction}
Quantum memories are essential elements to implement quantum logic, since the information must be preserved between gate operations. Different approaches to quantum memories are being studied, including NV centers in diamond, atomic gases, and single trapped atoms \cite{Simon:2010}. Superconducting circuits \cite{buluta2011,you2011atomic} are at the forefront in the race to realize the first quantum computers, because they exhibit flexibility, controllability and scalability. For this reason, quantum memories that can be easily integrated into superconducting circuits are also required. The realization of a quantum memory device, as well as of a quantum computer, is challenging because quantum states are fragile: the interaction with the environment causes decoherence. There are external, for example local electromagnetic signals, and intrinsic sources of decoherence. In circuit-QED, the main intrinsic source of decoherence are fluctuations in the critical-currents, charges, and magnetic-fluxes.
\\ \indent Superconducting circuits have allowed to achieve the ultrastrong coupling regime (USC) \cite{Niemczyk:2010gv,Tiefu2017,FornDiaz:2016bo}, where the light-matter interaction becomes comparable to the atomic and cavity frequency transitions ($\omega_q$ and $\omega_c$, respectively), reaching the coupling of $\lambda=1.34\,\omega_c$ \cite{yoshihara2017}. After a critical value of the coupling, $\lambda>\lambda_c$, with $\lambda_c=\sqrt{\omega_q\,\omega_c}/2$, the Dicke model predicts that a system of $N$ two-level atoms interacting with a single-cavity mode, in the thermodynamic limit ($N\to\infty$) and at zero temperature $(T=0)$, is characterized by a spontaneous polarization of the atoms and a spontaneous coherence of the cavity field. This situation can also be encountered in the finite-$N$ case \cite{emary2003,emary2004,ashhab2010}, in the limit of very strong coupling. 

Here, we consider a single two-level atom, $N=1$, interacting with a cavity mode in the USC regime. First, we derive a general master equation, valid for a large variety of hybrid quantum systems \cite{xiang2013hybrid} in the weak, strong, ultrastrong, and deep strong coupling regimes. Considering the two lowest eigenstates of our system, we show theoretically that if the coupling between the two-level atom and the cavity field is longitudinal and in the USC regime, the system is \textit{strongly protected against relaxation}. Vice versa, we prove that if the coupling is transverse and in the USC regime, then the system is \textit{protected against pure dephasing}. 

In the case of superconducting artificial atoms whose relaxation time is comparable to the pure dephasing time, taking advantage of this relaxation suppression in the USC regime, we prove that it is possible to apply the dynamical decoupling procedure  \cite{biercuk2011} to have full protection against decoherence. With the help of an auxiliary non-interacting atomic level, providing a suitable drive to the system, we show that a flying qubit that enters the cavity can be stored in our quantum memory device and retrieved afterwards. Moreover, we briefly analyze the case of artificial atoms transversally coupled to a cavity mode \cite{Nataf:2011ff,Kyaw:2015en}.

In this treatment we neglect the diamagnetic term $A^2$, which prevents the appearance of a superradiant phase, as the conditions of the no-go theorem can be overcome in circuit-QED \cite{nataf2010natcom,yoshihara2017}.
\begin{figure}[hbt]
  \includegraphics[scale=0.26]{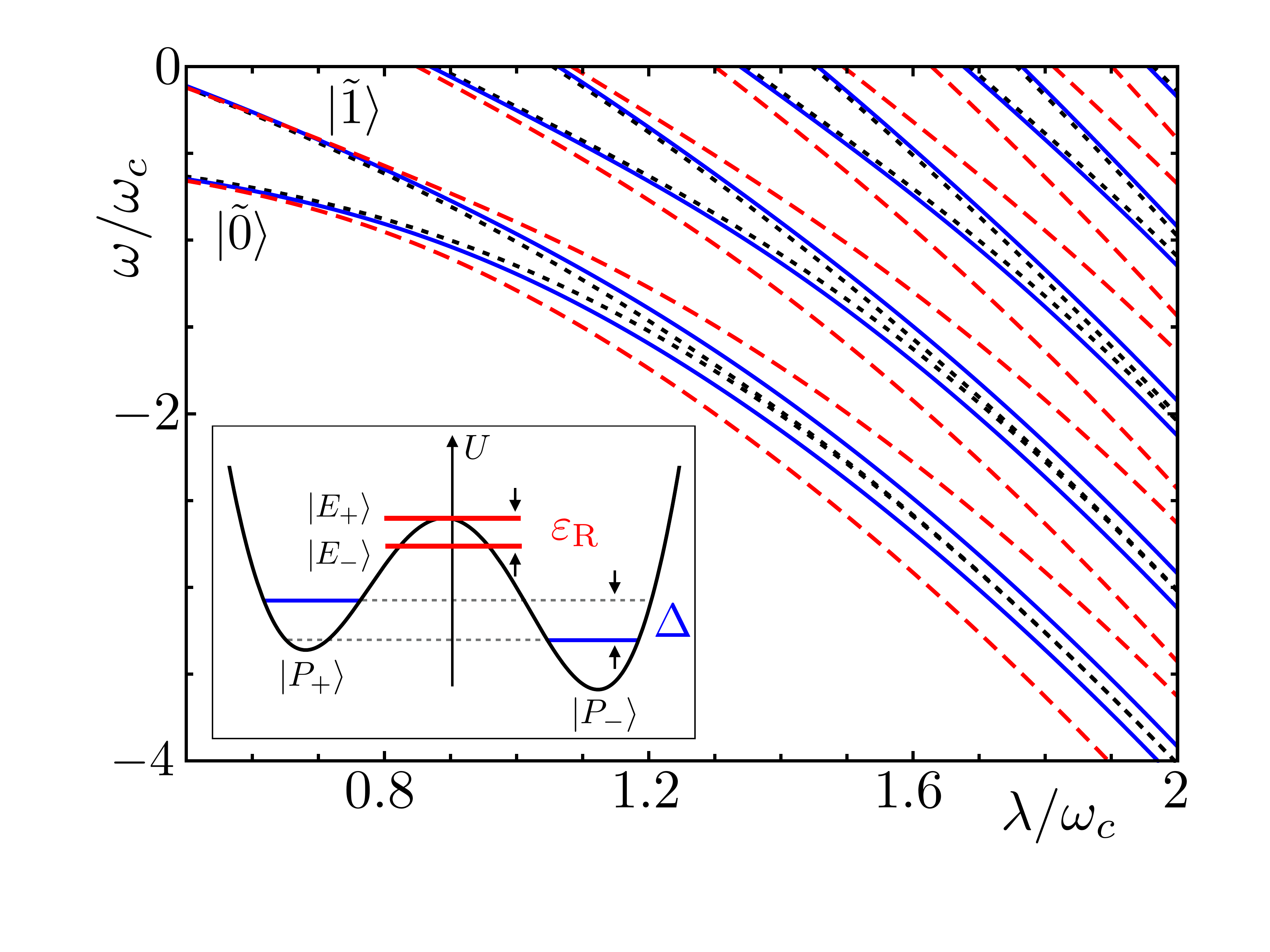}
  \caption{Energy levels for $\Delta=0$ (black dotted curves), $\Delta=0.2\,\omega_c$ (blue solid curves), and for $\Delta=0$ applying a constant field with $\Lambda=0.2\,\omega_c$ (red dashed curves). Here $\varepsilon=\omega_c=1$. Inset: graphical representation of the potential energy of the two-level system; each well is associated with a polarized state $\{\vert P_-\rangle, \vert P_+\rangle \}$.}\label{Fig1}
\end{figure}

\section{Model}
The Hamiltonian of a two-level system interacting with a cavity mode is $(\hbar=1)$\begin{equation}
\hat H=\omega_c \hat a^\dagger \hat a+\frac{\varepsilon}{2}\hat\sigma_z+\frac{\Delta}{2}\hat\sigma_x+\lambda\hat X\hat\sigma_x\, .
  \label{H}
\end{equation}
with $\hat a$ ($\hat a^\dagger$) the annihilation (creation) operator of the cavity mode with frequency $\omega_c$, $\hat X=\hat a+\hat a^\dagger$, and $\hat\sigma_j$ the Pauli matrices, with $j=\{x,y,z\}$. For a flux qubit, $\varepsilon$ and $\Delta$ correspond to the energy bias and the tunnel splitting between the persistent current states $\{\mid\, \downarrow\,\rangle,\mid\,\uparrow\,\rangle\}$ \cite{mooij1999}. 
We do not use the rotating wave approximation in the interaction term because the counterrotating terms are fundamental in the USC regime.  

For $\varepsilon=0$, the coupling is \textit{longitudinal} and the two lowest eigenstates $\{\vert \tilde 0\rangle,\vert \tilde 1\rangle\}$ are exactly the \textit{polarized} states $\vert P_-\rangle=\vert-\rangle\vert+\alpha\rangle$ and $\vert P_+\rangle=\vert +\rangle\vert-\alpha\rangle$, where $\vert\pm \rangle=1/\sqrt{2}(\,\mid\, \uparrow\,\rangle\,\pm\mid\,\downarrow\,\rangle)$, and $\vert\pm\alpha\rangle=\exp[\pm\alpha(\hat a^\dagger-a)]\vert 0\rangle$ are displaced Fock states \cite{Irish:2005ko}, with $\alpha=\lambda/\omega_c$. A proof of this is given in the Appendix \ref{PolStates}. In the subspace spanned by the polarized states $P=\{\vert P_-\rangle, \vert P_+\rangle \}$, $\hat H$ can be written, for $\varepsilon < \omega_c$,
\begin{equation}
  \hat H_P=\frac{\Delta}{2}\hat\sigma_z+\frac{\varepsilon_{\rm R}}{2}\hat\sigma_x\, ,
  \label{Hp-p+}
\end{equation}
with $\varepsilon_{\rm R}=\varepsilon\langle+\alpha\vert-\alpha\rangle$. Equation\,(\ref{Hp-p+}) describes a two-state system, see inset in Fig.\,\ref{Fig1}, characterized by a double-well potential with detuning parameter $\Delta$ and depth proportional to the overlap of the two displaced states. The kinetic contribution $(\varepsilon_{\rm R}/2)\hat\sigma_x$ mixes the states $P$ associated with the two minima of the potential wells.

For $\Delta=0$, the coupling is \textit{transverse} and the two lowest eigenstates $\{\vert \tilde 0\rangle,\vert \tilde 1\rangle\}$ converge, for $\lambda>\lambda_c$, to the \textit{entangled} states $\vert E_-\rangle=(\vert P_+\rangle-\vert P_-\rangle)/\sqrt{2}$ and $\vert E_+\rangle=(\vert P_+\rangle+\vert P_-\rangle)/\sqrt{2}$. In this case, as
\begin{equation}
\langle+\alpha\vert-\alpha\rangle=\exp\{-2\vert\lambda/\omega_c\vert^2\}\,,
\label{+alpha-alpha}
\end{equation}
the energy difference between the eigenstates, $\omega_{\tilde 1}-\omega_{\tilde 0}=\varepsilon_{\rm R}$, converges exponentially to zero with $\lambda$ (vacuum quasi-degeneracy), see Fig.\,\ref{Fig1} and Ref.\,\cite{nataf2010}. The system described by $\hat H$ does not conserve the number of excitations, $\mathcal N=a^\dagger a+\vert e\rangle\langle e\vert$, with $\vert e\rangle$ being the excited state of the two-level system, but for $\Delta=0$ has $\mathbb{Z}_2$ symmetry and it conserves the parity of the number of excitations \cite{Braak:2011,Schiro:2012}.

For $\Delta\neq 0$, the parity symmetry is broken \cite{garziano2014vac,Garziano:2015dv,Garziano:2016jy}. As $\varepsilon_R$ converges exponentially to zero with $\lambda$, the first two eigenstates of $\hat H$ converge exponentially to the polarized states $P$, and the energy splitting between the first two eigenstates converge to $\Delta$, see Eq.\,(\ref{Hp-p+}) and Fig.\,\ref{Fig1}.
For $\Delta=0$, it is also possible to break the $\mathbb{Z}_2$ parity symmetry, and have the polarized states $P$, applying to the cavity the constant field $-\Lambda/2\hat X$. In this case, the energy splitting between the first two eigenstates is a function of the coupling $\lambda$; indeed, $\omega_{\tilde 1}-\omega_{\tilde 0}=2\Lambda\lambda/\omega_c$, see Fig.\,\ref{Fig1} and Appendix \ref{casedelta0}.

\section{Master equation and coherence rate}
The dynamics of a generic open quantum system $S$, with Hamiltonian $\hat H_{\rm S}$ and eigenstates $\vert m\rangle$, is affected by the interaction with its environment $B$, described by a bath of harmonic oscillators. Relaxation and pure dephasing must be studied in the basis that diagonalizes $\hat H_{\rm S}$. 
The fluctuations that induce decoherence originate from the different channels that connect the system to its environment. For a single two-level system strongly coupled to a cavity field these channels are $\mathcal{S}=\{\hat\sigma_x, \hat\sigma_y, \hat\sigma_z, \hat X, \hat Y\}$, with $\hat Y=i(\hat a-\hat a^\dagger)$. 
In the interaction picture, the operators $\hat{S}^{(k)}\in\mathcal{S}$ can be written as
\begin{equation}
  \hat{S}^{(k)}\left(t\right)\,=\hat{S}^{(k)}_{+}\left(t\right)+\hat{S}^{(k)}_{-}\left(t\right)+\hat{S}^{(k)}_z\,,
\end{equation}
with 
\begin{equation}
  \hat{S}^{(k)}_{-}(t)=\sum_{m,n>m}s^{(k)}_{mn}\,\vert m\rangle\langle n\vert\, e^{-i\omega_{nm}t}\,,
\end{equation}
\begin{equation}
  \hat{S}^{(k)}_z=\sum_m s^{(k)}_{mm}\,\vert m\rangle\langle m\vert 
\end{equation}
and $\hat S^{(k)}_+=(\hat S^{(k)}_-)^\dagger$; this in analogy with $\hat\sigma_+$, $\hat\sigma_-$ and $\hat\sigma_z$, for a two-state system \cite{Carmichael1991}, while $s^{(k)}_{mn}=\langle m\vert\hat{S}^{(k)}\vert n\rangle$ and  $\omega_{mn}=\omega_{m}-\omega_{n}$.
The interaction of the environment with $\hat{S}^{(k)}_z$ affects the eigenvalues of the system, and involves the randomization of the relative phase between the system eigenstates. The interaction of the environment with $\hat{S}^{(k)}_x=\hat{S}^{(k)}_{+}+\hat{S}^{(k)}_{-}$ induces transitions between different eigenstates.
With this formulation, we have derived a master equation in the Born-Markov approximation valid for generic hybrid-quantum systems \cite{Beaudoin:2011ip}, at $T=0$,
\begin{eqnarray}
  \dot{\hat\rho} &=&	-i\left[\hat {H}_{\rm S},\hat\rho\right]+\sum_k\sum_{m,\,n>m}\Gamma^{(k)}_{mn}\mathcal{D}\left[\vert m\rangle\langle n\vert\right]\hat\rho\\ 
  &&+\sum_k\gamma^{(k)}_{\rm \varphi}\mathcal D\left[\hat{S}^{(k)}_z \right]\hat\rho\,,\nonumber
\end{eqnarray}
where $\mathcal{D}[\hat O]\hat\rho=(2\hat O\hat\rho\,\hat O^\dagger-\hat O^\dagger\hat O\hat\rho-\hat\rho\,\hat O^\dagger\hat O)/2$ is the Lindblad superoperator. The sum over $k$ takes into account all the channels $\hat{S}^{(k)}\in\mathcal{S}$. $\Gamma^{(k)}_{mn}=\gamma^{(k)}(\omega_{mn})\vert s^{(k)}_{mn}\vert^2$  are the transition rates from level $n$ to level $m$, $\gamma^{(k)}(\omega_{mn})$ are proportional to the noise spectra. Expanding the last term in the above master equation, allows to prove that the pure dephasing rate is $\gamma^{(k)}_{\rm \varphi}\vert s^{(k)}_{mm}-s^{(k)}_{nn}\vert^2/4$.
Using only the lowest two eigenstates of $\hat H_{\rm S}$, the master equation can be written in the form 
\begin{equation}
  \dot{\hat\rho}=-i\left[\hat {H},\hat\rho\right]+\sum_k\Gamma^{(k)}\mathcal{D}\left[\hat\sigma_-\right]\hat\rho+\gamma^{(k)}_{\rm \varphi}\mathcal D\left[\hat{S}^{(k)}_z \right]\hat\rho\,,
\end{equation}
where $\hat\sigma_-$ is the lowering operator. In the weak- or strong-coupling regime, it corresponds to the classical master equation in the Lindblad form for a two-state system.  For a complete derivation of the master equation, see Appendix \ref{MEgens}.
\section{Analysis}
As shown above, if the coupling is transverse, in the USC regime the two lowest eigenstates converge to the entangled states $E=\{\vert E_-\rangle,\vert E_+\rangle\}$ as a function of the coupling $\lambda$. If the coupling is longitudinal, the two lowest eigenstates are the polarized states $P$.  Moreover, we proved that the relaxation of the population is proportional to $\vert s^{(k)}_{mn}\vert^2$ and the pure dephasing to $\vert s^{(k)}_{mm}-s^{(k)}_{nn}\vert^2/4$; we call these two quantities \textit{sensitivity} to longitudinal relaxation and to pure dephasing, respectively. In  Table\,\ref{Table1} we report the values of 
\begin{subequations}
\begin{eqnarray}
	& S_{\rm R}(C)=\vert\langle C_+\vert\hat{S}\vert C_-\rangle\vert &  \\
	& S_{\rm D}(C)=\vert\langle C_+\vert\hat{S}\vert C_+\rangle-\langle C_-\vert\hat{S}\vert C_-\rangle\vert/2\,, &
\end{eqnarray}
\end{subequations}
calculated for every channel $\hat S$ in $\mathcal S$, and $C$ is $E$ or $P$. As $\langle+\alpha\vert-\alpha\rangle$ converges exponentially to zero with $\lambda$, see Eq.\,(\ref{+alpha-alpha}), if the coupling is \textit{longitudinal}, there is protection against \textit{relaxation}; if the coupling is \textit{transverse}, there is protection against \textit{pure dephasing}. The suppression of the relaxation can be easily understood considering that, increasing the coupling $\lambda$, increases the displacement and the depth of the two minima associated with the double well represented in the inset of Fig.\,\ref{Fig1}. The sensitivity to the relaxation $\vert s^{(k)}_{mn}\vert^2$ is connected to Fermi's golden rule for first-order transitions. Considering the polarized states $P$, the suppression of the longitudinal relaxation rates holds for every order. This is because every other intermediate path between the $P$ states, through higher states, involves always atomic and photonic coherent states with opposite signs.

When the coupling is transverse, the suppression of the pure dephasing is given by the presence of the photonic coherent states $\vert \pm\alpha\rangle$, which suppress the noise coming from the $\hat\sigma_z$ and $\hat\sigma_y$ channels \cite{Nataf:2011ff}, while for the other channels the system is in a \textquotedblleft sweet spot\textquotedblright. For this reason, this suppression holds only to first order. Furthermore, approaching the vacuum degeneracy, fluctuations in $\Delta$ become relevant and they drive the entangled states $E$ to the polarized states $P$ (spontaneous breaking of the parity symmetry \cite{garziano2014vac}). This will be further explained in Section \ref{ssectpd}.
 \begin{table}[t]
\caption{\label{Table1}%
Values of $S_{\rm R}(E)$, $S_{\rm D}(E)$, $S_{\rm R}(P)$ and $S_{\rm D}(P)$  calculated for every channel in $\mathcal S$. }
\begin{ruledtabular}
\begin{tabular}{ccccc}
$\hat S$ & $S_{\rm R}(E)$ &
\multicolumn{1}{c}{\textrm{$S_{\rm D}(E)$}}&
\multicolumn{1}{c}{\textrm{$S_{\rm R}(P)$}}&
\multicolumn{1}{c}{\textrm{$S_{\rm D}(P)$}}\\
\hline
$\hat\sigma_x$&\mbox{1} &0&0& 1 \\
$\hat\sigma_y$&$i\langle-\alpha\vert+\alpha\rangle$& 0 &$i\langle-\alpha\vert+\alpha\rangle$& 0 \\
$\hat\sigma_z$& 0 & $\langle+\alpha\vert-\alpha\rangle$ & $\langle-\alpha\vert+\alpha\rangle$ & 0\\
$\hat X$& $2\alpha$ & 0 & 0 & $2\alpha$ \\
$\hat Y$& 0 & 0 & 0 & 0 \\
\end{tabular}
\end{ruledtabular}
\end{table}

\section{Dynamical decoupling}
The dynamical decoupling (DD) method \cite{Viola:1998ky} consists of a sequence of $\pi$-pulses that average away the effect of the environment on a two-state system.  To protect from pure dephasing, the DD method uses a sequence of $\hat\sigma_x$ or $\hat\sigma_y$ pulses. If we rotate the $\hat\sigma_z$ and $\hat\sigma_y$ operators in the basis given by the states $P$, we find that $\hat R\,\hat\sigma_z\hat R^{-1}=\beta^{-1}\hat\sigma_x$ and $\hat R\,\hat\sigma_y\hat R^{-1}=\beta^{-1}\hat\sigma_y$, with $\beta^{-1}=\langle+\alpha\vert-\alpha\rangle$. Therefore,  $\hat\sigma_z$ and $\hat\sigma_y$ pulses in the bare atom basis correspond to $\hat\sigma_x$ and $\hat\sigma_y$ pulses attenuated by the $\beta^{-1}$ factor in the basis given by the states $P$. To compensate the reduction, the amplitude of the pulses must be multiplied by a factor $\beta$.
When the direction of the coupling is not exactly longitudinal, the convergence of the lowest eigenstates to the polarized states $P$ is exponential with respect to the coupling; thus, the $\hat\sigma_z$ operator in the free-atom basis is not exactly the $\hat\sigma_x$ operator in the reduced eigenbasis of $\hat H$. 
Instead, there are no problems with the $\hat\sigma_y$ operator of the bare atom, because it corresponds exactly to $\beta^{-1}\hat\sigma_y$ in the reduced dressed basis.
 
\begin{figure}[hbt]
  \includegraphics[scale=0.64]{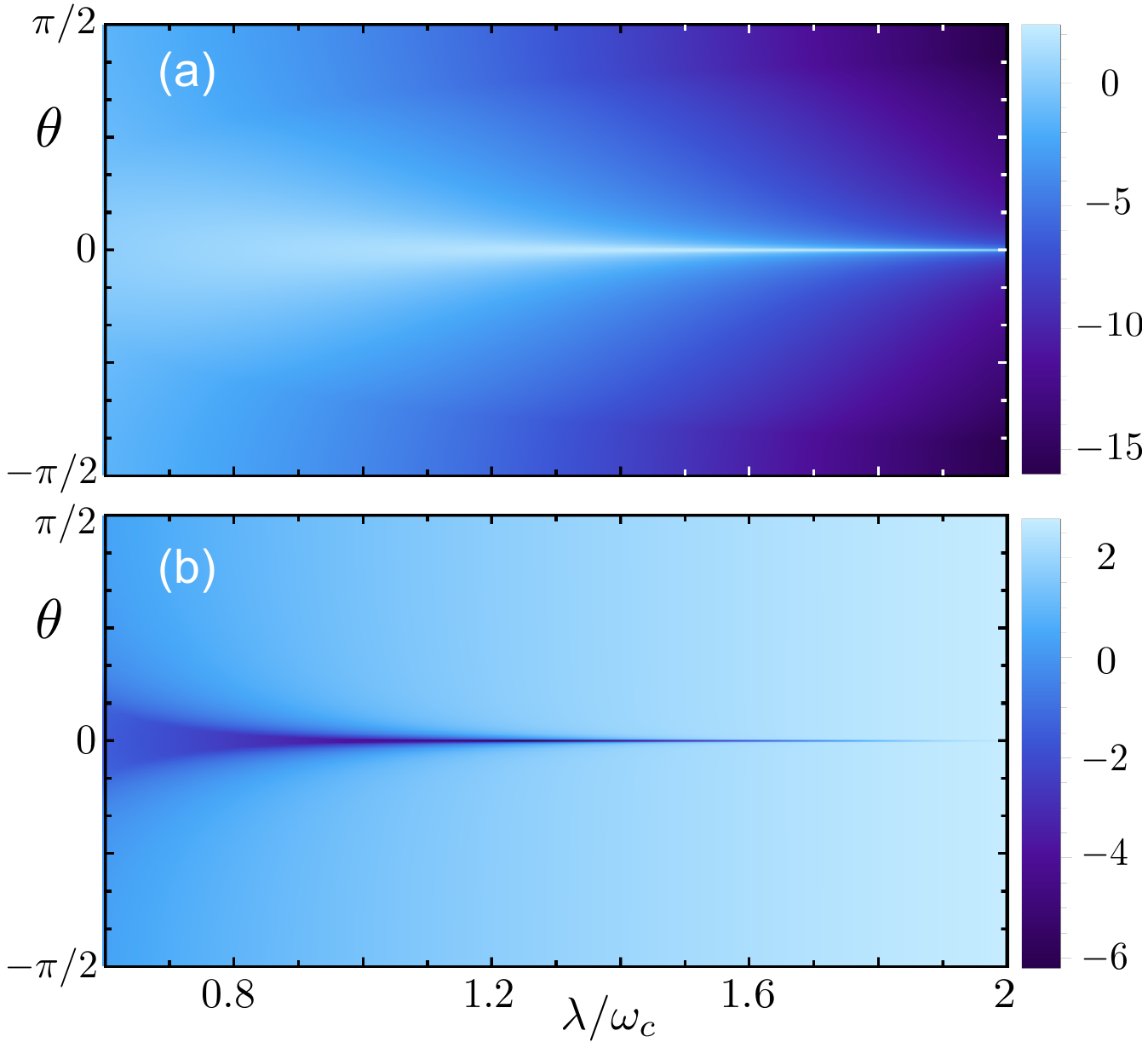}
  \caption{(a) Contour plot in a logarithmic scale (vertical bar on the right) of the maximum sensitivity to relaxation, $\rm{max}\{\vert s^{(k)}_{\tilde 0\tilde 1}\vert^2:\hat S^{(k)}\in \mathcal{S} \}$, versus the normalized coupling $\lambda/\omega_c$ and the angle $\theta$. (b) Contour plot in a logarithmic scale of the maximum sensitivity to pure dephasing, $\rm{max}\{\vert s^{(k)}_{\tilde 1\tilde 1}-s^{(k)}_{\tilde 0\tilde 0}\vert^2/4:\hat S^{(k)}\in\mathcal S\}$, versus the normalized coupling $\lambda/\omega_c$ and $\theta$. Here, $\omega_q=0.2\,\omega_c$, and $\omega_c=1$.}
  \label{Fig2}
\end{figure}

\section{Proposal}
\subsection{$T_1<T_\varphi$ or $T_1\sim T_\varphi$}
This proposal is applicable to supeconducting qubits whose relaxation time $T_1$ is lower than the pure dephasing time $T_\varphi$ or comparable, i.e. flux qubits.
If we consider the polarized states $P$ as a quantum memory device and if we prepare it in an arbitrary superposition, we can preserve coherence. Indeed, our quantum memory device is naturally protected from population relaxation. To protect it from pure dephasing, we apply DD \cite{Bylander:2011}.
We consider $\hat H$ in Eq.\,(\ref{H}) with $\Delta\neq 0$. In order to have the second excited states far apart in energy, we need $\vert\Delta\vert <0.5\,\omega_c$. 
The longitudinal relaxation suppression behaves as $\vert\langle+\alpha\vert-\alpha\rangle\vert^{2}=\exp\{-4N(\lambda/\omega_c)^2\}$; increasing the coupling $\lambda$ or the number $N$ of atoms increases exponentially the decay time of the longitudinal relaxation. However, the contribution of the $\hat X$ channel to pure dephasing increases quadratically with $\lambda/\omega_c$. This does not affect the coherence time of our system; indeed, superconducting harmonic oscillators generally have higher quality factors than superconducting qubits. 
It is convenient to write $\hat H$ in Eq.\,(\ref{H}) in the basis that diagonalizes the atomic two-level system $\{\vert g\rangle,\vert e\rangle\}$,
\begin{equation}
  \hat {H}'=\omega_c \hat a^\dagger \hat a+\frac{\omega_q}{2}\hat\sigma_z+\lambda\hat X\left(\cos\theta\,\hat\sigma_x+\sin\theta\,\hat\sigma_z\right)\,,
  \label{Hniem}
\end{equation}
with $\theta=\arctan(\Delta/\varepsilon)$ and $\omega_q=\sqrt{\varepsilon^2+\Delta^2}$. 
Using Eq.\,(\ref{Hniem}), in Fig.\,\ref{Fig2}(a) we show  the numerically calculated sensitivity, $\rm{max}\{\vert s^{(k)}_{\tilde 0\tilde 1}\vert^2:\hat S^{(k)}\in \mathcal{S} \}$, to the longitudinal relaxation as a function of the normalized coupling $\lambda/\omega_c$ and of the angle $\theta$. For large values of $\lambda/\omega_c$ and for $\theta\neq 0$, there is a strong suppression of the relaxation rate: it is maximum when the coupling is entirely longitudinal, $\theta =\pi/2$. 
For $\lambda/\omega_c=1.3$, $\theta=\pi/2$ and $\omega_q=0.2\,\omega_c$, the longitudinal relaxation rate is reduced by a factor $\approx 10^{-3}$, meanwhile the contribution of the cavity field to the pure dephasing rate increases only by $6.76$. 
Moreover, for one two-state system affected by $1/f$ noise, the DD can achieve up to $10^3$-fold enhancement of the pure dephasing time $T_\varphi$, applying $1000$ equally spaced $\pi$-pulses (see Appendix \ref{appDD}). Using this proposal with these parameters, it is possible to \textit{increase the coherence time of a superconducting two-level atom up to $10^3$ times}.
\subsection{$T_1\gg T_\varphi$}\label{ssectpd}
Figure \ref{Fig2}(b) shows the numerically calculated maximum sensitivity to pure dephasing, $\rm{max}\{\vert s^{(k)}_{\tilde 1\tilde 1}-s^{(k)}_{\tilde 0\tilde 0}\vert^2/4:\hat S^{(k)}\in\mathcal S\}$, as a function of $\lambda/\omega_c$ and $\theta$. For large values of $\lambda/\omega_c$, the strong suppression of the pure dephasing rate is confined to a region (dark blue) that exponentially converges to zero for increasing $\lambda$; only in this region the entangled states exist. In Fig.\,\ref{Fig2}(b), for $\Delta=0\, (\theta=0)$, it is clear that, for a large value of the coupling $\lambda$, fluctuations in $\Delta$ (or in $\theta$) drive the entangled states $E$ (dark blue region) to the polarized states $P$ (light blue region). 
Superconducting qubits whose relaxation time $T_1$ is much greater than the pure dephasing time $T_\varphi$, i.e. fluxonium \cite{Pop:2014gg}, can take advantage of the suppression of the pure dephasing. For $\lambda/\omega_c=0.8$, $\theta=0$ and $\omega_q=0.5\,\omega_c$, the pure dephasing rate is reduced by a factor $\approx 7\times 10^{-2}$; meanwhile the contribution of the cavity field to the longitudinal relaxation rate increases only by $2.47$.

\section{Protocol}
\begin{figure}[hbt]
  \includegraphics[scale=0.65]{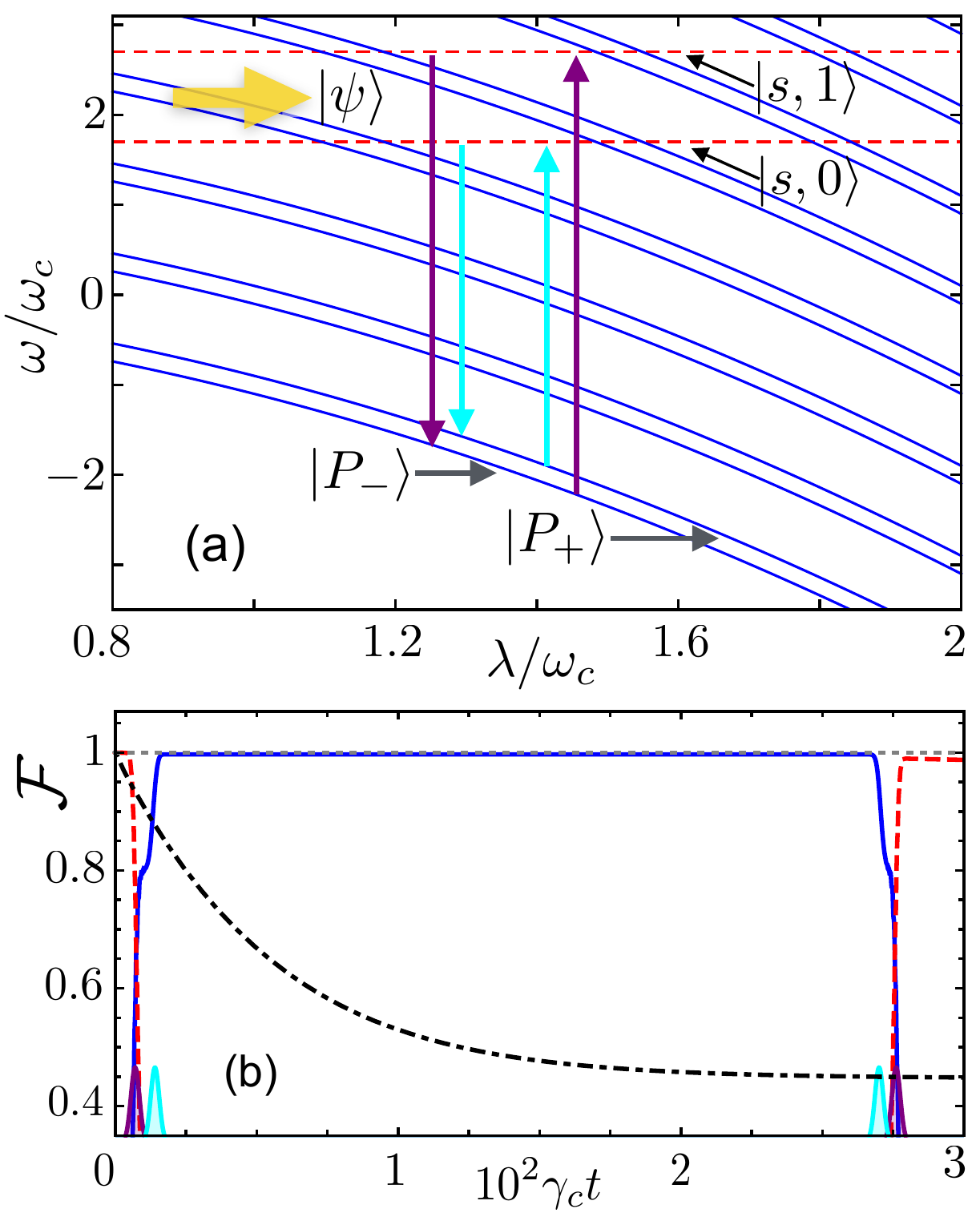}
  \caption{Two-level system ultrastrongly coupled to a cavity mode and an auxiliary non-interacting level $s$. (a) Energy levels of $\hat H_{\rm tot}$ versus the normalized coupling $\lambda/\omega_c$. The blue solid curves concern the interacting part; the red dashed horizontal lines concern the non-interacting part. (b) Time evolution of the fidelity $\mathcal F$ between the initial state $\vert\psi\rangle$ and $\vert\Psi_{\rm s}(t)\rangle$ (red dashed curve), $\vert\Psi_{\rm P}(t)\rangle$ (blue solid curve), and the atomic state in the non-interacting case (black dotted-dashed curve). Here, $\omega_c=1$, $\varepsilon=0.01\,\omega_c$, $\Delta=0.2\,\omega_c$, $\lambda=1.3\,\omega_c$, $\omega_{\rm s}=1.7\,\omega_c$. The cavity and $\vert s\rangle\to\vert e\rangle$ relaxation rates are $\gamma_c=\gamma_{se}=10^{-5}\omega_c$.}
  \label{Fig3}
\end{figure}

Now we propose a protocol to write-in and read-out the quantum information encoded in a Fock state $\vert\psi\rangle=a\vert 0\rangle + b\vert 1\rangle$. 
We consider an auxiliary atomic state $\vert s\rangle$ decoupled from the cavity field, and with higher energy $\omega_{\rm s}$ respect to the two-level system $\{\vert g\rangle,\vert e\rangle\}$ \cite{Liu:2005eo,deppe2008two}. Figure \ref{Fig3}(a) shows the eigenvalues of the Hamiltonian of the total system, $\hat H_{\rm tot}=\hat {H}'+\omega_{\rm s}\vert s\rangle\langle s\vert$, versus the coupling $\lambda/\omega_c$. The blue solid curves concern $\hat {H}'$, the red dashed equally-spaced lines the auxiliary level $\vert s\rangle$ and these count the number of photons in the cavity \cite{Stassi:2013prl}.
We prepare the atom in the state $\vert s\rangle$ sending a $\pi$-pulse resonant with the transition frequency between the ground  $\vert P_-\rangle$ and $\vert s, 0\rangle$ states \cite{di2016cutting}. When the qubit with an unknown quantum state $\vert\psi\rangle$ enters  the cavity, the state becomes $\vert\Psi_{\rm s} \rangle=\vert s\rangle\otimes(a\vert 0\rangle+b\vert 1\rangle)=a\vert s, 0\rangle+b\vert s, 1\rangle$. Immediately after, we send two $\pi$-pulses: $p_1$ resonant with the transition $\vert s,1\rangle\to\vert P_- \rangle$ and $p_2$ resonant with the transition $\vert s, 0 \rangle\to\vert P_+\rangle$. Hereafter, we apply DD to preserve the transverse relaxation rate; meanwhile the quantum memory device is naturally protected from the longitudinal relaxation. To restore the quantum information we reverse the storage process. Figure \ref{Fig3}(b) shows the time evolution of the fidelity $\mathcal F$ between the initial state $\vert\psi\rangle$ and the states $\vert\Psi_{\rm s}(t)\rangle=a_{\rm s}\vert s,0\rangle+b_{\rm s}\vert s,1\rangle$ and $\vert\Psi_{\rm P}(t)\rangle=a_+\vert P_+\rangle+b_-\vert P_-\rangle$ in the rotating frame, this is calculated using the above master equation for $\lambda=1.3\,\omega_c$. The standard decay rates are assumed to be the same for every channel of the two-level artificial atom $\{\vert g\rangle,\vert e\rangle\}$, $\gamma^{(k)}=10^{-3}\omega_c$. For the pure dephasing rates, we choose $\gamma^{(k)}_\varphi=10^{-3}\gamma^{(k)}$, since we apply DD. The pulses are described by $\hat H_{\rm p_1}=\epsilon(t)\cos(\omega_{mn} t)(\hat\sigma_{\rm gs}+\hat\sigma_{\rm gs}^\dagger)/\langle m\vert\hat\sigma_{\rm gs}\vert n\rangle$ and $\hat H_{\rm p_2}=\epsilon(t)\cos(\omega_{mn} t)(\hat\sigma_{\rm es}+\hat\sigma_{\rm es}^\dagger)/\langle m\vert\hat\sigma_{\rm es}\vert n\rangle$, where $\hat\sigma_{\rm gs}=\vert g \rangle\langle s\vert$, $\hat\sigma_{\rm es}=\vert e \rangle\langle s\vert$, and $\epsilon(t)$ is a Gaussian envelope. 
At time $t=0$, the states  $\vert s,0\rangle$ and $\vert s,1\rangle$ are prepared, so that $a_{\rm s}^2=0.8$ and $b_{\rm s}^2=0.2$. As shown in Fig.\,\ref{Fig3}(b), at times $\gamma_{c} t_1=7\times 10^{-4}$ and $\gamma_{c} t_2=14\times 10^{-4}$, we apply the pulses $p_{1}$ and $p_{2}$, respectively. Now the populations and the coherence are completely transferred to the polarized states $P$, and the qubit is stored. Later, at $\gamma_c t_3=2.7\times 10^{-2}$ and $\gamma_c t_4=2.76\times 10^{-2}$, two pulses equal to the previous ones restore the qubit $\vert\psi\rangle$ into the cavity. 
As a comparison, we have calculated the fidelity (black curve) between $\vert\psi\rangle$ and the state of a two-level artificial atom prepared at $t=0$ in the same superposition as $\vert\psi\rangle$, but  interacting ordinarily with the cavity field, $\lambda/\omega_c\ll 0.1$, and now without DD (free decay). This fidelity converges to its minimum value much faster than the one calculated for the polarized states, which is not significantly affected by decoherence in the temporal range shown in [Fig.\,\ref{Fig3}(b)].

\section{Conclusions}
We propose a quantum memory device composed of the lowest two eigenstates of a system made of a two-level atom and a cavity mode interacting in the USC regime when the parity symmetry of the Rabi Hamiltonian is broken. 
Making use of an auxiliary non-interacting level, we store and retrieve the quantum information. 
For parameters adopted in the simulation, it is possible to improve the coherence time of a superconducting two-state atom up to $10^3$ times. For instance, the coherence time of a flux qubit longitudinally coupled to a cavity mode \cite{Billangeon:2015dn,Richer:2016hu,wang2016multiple}, at the optimal point, can be extended from $10\, \mu \rm s$ to over $0.01$ seconds \cite{Stern:2014hh}. Instead, in the case of unbroken parity symmetry, the coherence time of a fluxonium, with applied magnetic flux $\Phi_{\rm ext}=0.5\,\Phi_0$, inductively coupled to a cavity mode, can be extended from $14\,\mu \rm s$ to $0.2\, \rm ms$ \cite{Pop:2014gg}. This is a remarkable result for many groups working with superconducting circuits. Similar approaches can be applied to other types of qubits.


\appendix
\section{Polarized States $\{\vert P_-\rangle, \vert P_+\rangle \}$}\label{PolStates}

In this Appendix, we prove that when the coupling between a two-level system and a cavity mode is longitudinal, the two lowest eigenstates are the \textit{polarized} states $\vert P_-\rangle=\,\vert-\rangle\vert+\alpha\rangle$ and $\vert P_+\rangle=\,\vert+\rangle\vert-\alpha\rangle$, where $\vert\pm \rangle=1/\sqrt{2}(\,\vert \uparrow\,\rangle\pm \,\vert\downarrow\,\rangle)$, $\{\mid\, \downarrow\,\rangle,\mid\,\uparrow\,\rangle\}$ are, for example, persistent current states in the case of a flux qubit, and $\vert\pm\alpha\rangle=\exp[\pm\alpha(\hat a^\dagger-a)]\vert 0\rangle$ are displaced Fock states, with $\alpha=\lambda/\omega_c$.

\subsection{Case: $\Delta\neq 0$}
Let us start with the Hamiltonian of a two-level system interacting \textit{longitudinally} with a cavity mode 
\begin{equation}
\hat H=\omega_c\, \hat a^\dagger \hat a+\frac{\Delta}{2}\hat\sigma_x+\lambda\hat X\hat\sigma_x\,.
  \label{HApp}
\end{equation}

Replacing $\hat{\sigma}_x$ by its eigenvalue $m=\pm1$, we can write

\begin{equation}
\hat H=\omega_c\, \hat a^\dagger \hat a+m\left(\frac{\Delta}{2}+\lambda\hat X\right)\,.
\label{Hoscll}
\end{equation}

The transformation $\hat{a}=\hat{b}- m\lambda /\omega_c$, which preserves the commutation relation between $\hat{a}$ and  $\hat{a}^{\dagger}$, $[\hat{b},\hat{b}^\dagger]=1$, diagonalizes $\hat H$

\begin{equation}
  \hat{H}=\omega_c\,\hat{b}^\dagger\hat{b}-\frac{\lambda^2 m^2}{\omega_c}+\frac{\Delta}{2}m  \,.
  \label{Hdiaglinf1}
\end{equation}
This is the Hamiltonian of a displaced harmonic oscillator. Applying the operator $\hat b=\hat a +m\alpha$, with $\alpha=\lambda/\omega_c$, to the ground state $\vert 0_{m}\rangle$ of the oscillator given by Eq.\,(\ref{Hdiaglinf1}), gives $\hat a\vert 0_m\rangle=-m\alpha\vert 0_m\rangle$. We now see that $\vert -m\alpha\rangle=\vert 0_m\rangle$ is a coherent state with eigenenergy
\begin{equation}
  \omega_m=-\frac{\lambda^2m^2}{\omega_c}+m\frac{\Delta}{2}\,.
\end{equation}

Therefore, the two lowest eigenstates of the Hamiltonian $\hat{H}$ in Eq.\,(\ref{HApp}) are the two states $\vert P_{-}\rangle=\vert -\rangle\vert+\alpha \rangle$ and $\vert P_{+}\rangle=\vert +\rangle\vert-\alpha \rangle$, with eigenvalues $\omega_\pm=-\lambda^2m^2/\omega_c\pm\Delta/2$. The energy splitting between the eigenstates $\vert P_{-}\rangle$ and $\vert P_{+}\rangle$ is $\omega_+-\omega_-=\Delta$. The number of photons contained in each state is $n=\vert\alpha\vert^2=\lambda^2/\omega_c^2$.
\subsection{Case: $\Delta= 0$}\label{casedelta0}
The polarized states can be generated also substituting in Eq.\,(\ref{HApp}) the term $\Delta\hat\sigma_x/2$ with the field $-\Lambda\left(a+a^{\dagger}\right)/2$
\begin{equation}
  \hat{H}=\omega_c \hat{a}^\dagger \hat{a}-\frac{\Lambda}{2}\hat{X}+\lambda \hat{X}\hat{\sigma}_x\,.
  \label{HlambdainfCav}
\end{equation}
Following the same procedure as in the previous case, we can write
\begin{equation}
  \hat{H}=\omega_c \hat{a}^\dagger \hat{a}+\left(\hat{a}+\hat{a}^\dagger\right)\left(m\lambda-\frac{\Lambda}{2}\right),
\end{equation}
that can be diagonalized by the transformation $\hat{a}=\hat{b}-(m\lambda-\Lambda/2)/\omega_c$,
\begin{equation}
  \hat{H}=\omega_c\hat{b}^\dagger\hat{b}-\frac{\left(m\lambda-\frac{\Lambda}{2}\right)^2}{\omega_c} \,.
  \label{Hdiaglinf2}
\end{equation}
Considering the two lowest eigenstates, the excited state is now $\vert P_{+}\rangle=\vert +\rangle\vert-\alpha \rangle$ with energy $\omega_+=-\left(\Lambda/2-\lambda\right)^2/\omega_c$ and the ground state is $\vert P_{-}\rangle=\vert -\rangle\vert+\alpha \rangle$ with energy $\omega_-=-\left(\Lambda/2+\lambda\right)^2/\omega_c$, and $-m\alpha=-(m\lambda-\Lambda/2)/\omega_c$. The energy difference between the excited and the ground state is $\omega_+-\omega_-=2\lambda\Lambda/\omega_c$.

\section{Master equation for a generic hybrid system}\label{MEgens}

The total Hamiltonian that describes a generic hybrid system interacting with the environment $B$ is 
\begin{equation}
  \hat H=\hat H_S+\hat H_B+\hat H_{SB}\,,
\end{equation}
where $\hat H_S$, $\hat H_B$ and $\hat H_{SB}$, are respectively the Hamiltonians of the system, bath, and system-bath interaction. Here, $\hat H_{SB}=\sum_k\hat H^{(k)}_{SB}$, where the sum is over all the channels $k$ that connect the system $S$ to the environment. For a single two-level system strongly coupled to a cavity field these channels are $\mathcal{S}=\{\hat\sigma_x, \hat\sigma_y, \hat\sigma_z, \hat X, \hat Y\}$, with $\hat Y=i(\hat a-\hat a^\dagger)$.
In the interaction picture we have
\begin{eqnarray}\label{SinAuto}
  \hat{S}^{(k)}\left(t\right)\;&=&\;\sum_{mn} s^{(k)}_{mn}\,\vert m\rangle\langle n\vert\, e^{i\omega_{mn}t}
  \;\\ \nonumber
  &=&\hat{S}^{(k)}_{+}\left(t\right)+\hat{S}^{(k)}_{-}\left(t\right)+\;\hat{S}^{(k)}_z\,,
\end{eqnarray}
with
\begin{equation}
  \hat{S}^{(k)}_{-}(t)=\sum_{m,n>m}s^{(k)}_{mn}\,\vert m\rangle\langle n\vert\, e^{-i\omega_{nm}t}\,,
\end{equation}
\begin{equation}
  \hat{S}^{(k)}_z=\sum_m s^{(k)}_{mm}\,\vert m\rangle\langle m\vert
\end{equation}
and $\hat S^{(k)}_+=(\hat S^{(k)}_-)^\dagger$, this in analogy with $\hat\sigma_+$, $\hat\sigma_-$ and $\hat\sigma_z$ for a two-state system \cite{Carmichael1991}, where $s^{(k)}_{mn}=\langle m\vert\hat{S}^{(k)}\vert n\rangle$ and $\omega_{mn}=\omega_{m}-\omega_{n}$.
 The interaction of the environment with $\hat{S}^{(k)}_z$ affects  the eigenstates of the system, and involves the randomization of the relative phase between the system eigenstates. The interaction of the environment with $\hat{S}^{(k)}_x=\hat{S}^{(k)}_{+}+\hat{S}^{(k)}_{-}$ induces transitions among different eigenstates. We use  the Born master equation in the interaction picture
\begin{equation}\label{BornMaster}
  \dot{\hat\rho}_I=-\frac{1}{\hbar^2}\sum_k\int_0^t d{t}'\,\rm{tr}_B\left\{\left[\hat H^{(k)}_{SB}\left(t\right),\left[\hat H^{(k)}_{SB}\left({t}'\right),\hat\rho_I\left({t}'\right)\hat B_0\right] \right]\right\}
\end{equation}
where $\hat B_0$ is the density operator of the bath at $t=0$.
\subsection{Relaxation}
Within the general formula for a system $S$ interacting with a bath $B$, described by a bath of harmonic oscillators, in the rotating wave approximation, the Hamiltonian $\hat H_{SB}$ is

\begin{equation}\label{HSB}
  \hat H_{SB}^{(k)}\left(t\right)=\hat{S}^{(k)}_{-}(t)\hat B^\dagger(t)+\hat{S}_{+}^{(k)}(t)\hat B(t)
\end{equation}

with $\hat B(t)=\sum_p\kappa\hat b_p e^{-i\nu_p t}$, where $\kappa$ is the coupling constant with the system operator $\hat S^{(k)}$. We assume that the bath variables are distributed in the uncorrelated thermal mixture of states. It is easy to prove that
\begin{eqnarray}\label{BathCorr}
  &&\langle \hat B(t)\hat B({t}')\rangle_B=0\,,\\ \nonumber
  &&\langle \hat B^\dagger (t)\hat B^\dagger ({t}')\rangle_B=0\,,\\ \nonumber
  &&\langle \hat B^\dagger(t)\hat B({t}')\rangle_B=\sum_p\kappa^2 \exp\{i\nu_p(t-{t}')\}\bar n(\nu_p,T)\,,\\ \nonumber
  &&\langle \hat B(t)\hat B^\dagger({t}')\rangle_B=\sum_p\kappa^2 \exp\{-i\nu_p(t-{t}')\}[1+\bar n(\nu_p,T)]\,, \nonumber
\end{eqnarray}
 where $\bar n=(\exp\{\frac{\hbar\nu_p}{k_B T}\}-1)^{-1}$, $k_B$ is the Boltzmann constant, and $T$ is the temperature. Using Eq.\,(\ref{HSB}) and the properties of the trace, substituting $\tau=t-{t}'$, Eq.\,(\ref{BornMaster}) in the Markov approximation becomes $(\hbar=1)$
 
 \begin{eqnarray}
  \dot{\hat\rho}_I &=&\\ 
  &&\sum_k\sum_{(m,\,n>m)}\sum_{({m}',\,{n}'>{m}')}s^{(k)}_{mn}s^{(k)}_{{n}'{m}'} \nonumber \\ 
  &\times & \Big[\Big.\left(\vert {n}' \rangle\langle {m}' \vert\rho_I\vert m \rangle\langle n \vert - \vert m \rangle\langle n\vert {n}' \rangle\langle {m}' \vert\rho_I\right)\nonumber\\ 
   &\times & e^{i\left(\omega_{{n}'{m}'}-\omega_{nm} \right)t}\int_0^t d\tau\,e^{-i\omega_{{n}'{m}'}\tau} \langle \hat B^\dagger(t)\hat B(t-\tau)\rangle_B \nonumber \\   
   &+& \left(\vert {m}' \rangle\langle {n}' \vert\rho_I\vert n \rangle\langle m \vert - \vert n \rangle\langle m\vert {m}' \rangle\langle {n}' \vert\rho_I\right)\nonumber \\
   &\times &  e^{i\left(\omega_{nm}-\omega_{{n}'{m}'} \right)t}\int_0^t d\tau\,e^{i\omega_{{n}'{m}'}\tau} \langle \hat B(t)\hat B^\dagger(t-\tau)\rangle_B \nonumber \\ 
   &+& \left(\vert n \rangle\langle m \vert\rho_I\vert {m}' \rangle\langle {n}' \vert - \rho_I\vert {m}' \rangle\langle {n}'\vert n \rangle\langle m \vert\right)\nonumber \\
   &\times &  e^{i\left(\omega_{nm}-\omega_{{n}'{m}'} \right)t}\int_0^t d\tau\,e^{i\omega_{{n}'{m}'}\tau} \langle \hat B^\dagger(t-\tau)\hat B(t)\rangle_B \nonumber \\ 
   &+& \left(\vert m \rangle\langle n \vert\rho_I\vert {n}' \rangle\langle {m}' \vert - \rho_I\vert {n}' \rangle\langle {m}'\vert m \rangle\langle n \vert\right)\nonumber \\
   &\times &  e^{i\left(\omega_{{n}'{m}'}-\omega_{nm} \right)t}\int_0^t d\tau\,e^{-i\omega_{{n}'{m}'}\tau} \langle \hat B(t-\tau)\hat B^\dagger(t)\rangle_B\Big.\Big]\,.\nonumber
  \end{eqnarray}
  Within the secular approximation, it follows that ${m}'=m$ and ${n}'=n$.
  We now extend the $\tau$ integration to infinity and in Eqs.\,(\ref{BathCorr}) we change the summation over $p$ to an integral, $\sum_p\to\int_0^\infty d\nu\,g_k(\nu)$, where $g_k(\nu)$ is the density of states of the bath associated to the operator $\hat S^{(k)}$, for example
\begin{eqnarray}
   &&\int_0^t d\tau\,e^{-i\omega_{nm}\tau} \langle \hat B^\dagger(t)\hat B(t-\tau)\rangle_B\to \\ \nonumber
   && \int_0^{\infty} d\nu\,g_k\left(\nu\right)\kappa^2\left(\nu\right)\bar{n}\left(\nu,T\right)\int_0^\infty d\tau\,e^{i\left(\nu-\omega_{nm}\right)\tau}\,.
\end{eqnarray}
The time integral is $\int_0^\infty d\tau\,e^{i\left(\nu-\omega_{nm}\right)\tau}=\pi\delta(\nu-\omega_{nm})+i\mathcal{P}/(\nu-\omega_{nm})$, where $\mathcal{P}$ indicates the Cauchy principal value. We omit here the contribution of the terms containing the Cauchy principal value $\mathcal{P}$, because these represent the Lamb-shift of the system Hamiltonian. We thus arrive to the expression
\begin{widetext}
\begin{eqnarray}
  \dot{\hat\rho}_I &=& \pi\sum_k\sum_{m,\,n>m}\vert s^{(k)}_{mn}\vert^2\kappa^2\left(\omega_{mn}\right)g_k\left(\omega_{mn}\right)\left\{ \big(2\vert n \rangle\langle m \vert\rho_I\vert m \rangle\langle n \vert -\vert m \rangle\langle n\vert n \rangle\langle m \vert\rho_I - \rho_I\vert m \rangle\langle n\vert n \rangle\langle m \vert\big)\bar{n}\left(\omega_{mn},T\right)\right. \nonumber \\
  &+& \left. \big(2\vert m \rangle\langle n \vert\rho_I\vert n \rangle\langle m \vert
  -\vert n \rangle\langle m\vert m \rangle\langle n \vert\rho_I - \rho_I\vert n \rangle\langle m\vert m \rangle\langle n \vert\big)\left[\bar{n}\left(\omega_{mn},T\right)+1\right]\right\}\,,
\end{eqnarray}
\end{widetext}
with $s^{(k)}_{nm}=(s^{(k)}_{mn})^*$.
Transforming back to the Schr\"{o}dinger picture, we obtain the master equation for a generic system in thermal equilibrium 
\begin{eqnarray}\label{Relax}
\dot{\hat\rho}\left(t\right) &=&	-i\left[\hat {H}_S,\hat\rho\right]\\
&+&\sum_k\sum_{m,\,n>m}\Gamma^{(k)}_{mn}\left\{\mathcal{D}\Big[\vert n\rangle\langle m\vert\Big]\hat\rho\left(t\right)\bar{n}\left(\omega_{mn},T\right) \right. \nonumber \\
&+&\mathcal{D}\Big[\vert m\rangle\langle n\vert\Big]\hat\rho\left(t\right)\left[\bar{n}\left(\omega_{mn},T\right)+1\right]\Big.\Big\} \nonumber
\end{eqnarray}
where $\Gamma^{(k)}_{mn}= 2\pi\vert s^{(k)}_{mn}\vert^2\kappa^2\left(\omega_{mn}\right)g_k\left(\omega_{mn}\right)$ is the transition rate from level $m$ to level $n$, and $\mathcal{D}[\hat O]\hat\rho=(2\hat O\hat\rho\,\hat O^\dagger-\hat O^\dagger\hat O\hat\rho-\hat\rho\,\hat O^\dagger\hat O)/2$.

\subsection{Pure dephasing}
A quantum model of the pure dephasing describes the interaction of the system with the environment in terms of virtual processes; 
the quanta of the bath with energy $\hbar\nu_{q}$ are scattered to quanta with energy $\hbar\nu_{p}$, leaving the states of the system unchanged. In the interaction picture we have
\begin{equation}
  \hat H^{(k)}_{SB}=\hat S_z^{(k)}\!\left(t\right)\hat B\!\left(t\right)
  \label{S19}
\end{equation}
with $\hat B\left(t\right)=\sum_{pq}\kappa\,\hat b^\dagger_p\,\hat b^{}_q\, e^{i\nu_{pq}t}$, where $\kappa$ is the coupling constant with the system. In the sum, terms with $p=q$ have nonzero thermal mean value and they will be included in $\hat H_S$, producing a shift in the Hamiltonian energies, so we will omit this contribution. Substituting Eq.\,(\ref{S19}) in the Born master equation Eq.\,(\ref{BornMaster}), with $\tau=t-{t}'$
\begin{eqnarray}
  \dot{\hat\rho}_I &=& \sum_k\sum_{m,{m}'}s^{(k)}_{m,m}s^{(k)}_{{m}',{m}'} \nonumber \\
  &\times & \Big[\Big(\vert {m}' \rangle\langle {m}' \vert\rho_I\vert m \rangle\langle m \vert-\vert m \rangle\langle m \vert {m}' \rangle\langle {m}' \vert\rho_I \big.\Big) \nonumber \\
  &\times & \int_0^t d\tau\langle \hat B\left(t\right)\hat B\left(t-\tau\right)\rangle_B\\  \nonumber
   &+&\Big(\vert m \rangle\langle m \vert\rho_I\vert {m}' \rangle\langle {m}' \vert-\rho_I\vert {m}' \rangle\langle {m}' \vert m \rangle\langle m \vert \Big)\big.\nonumber \\
   &\times & \int_0^t d\tau\langle \hat B\left(t-\tau\right)\hat B\left(t\right)\rangle_B \Big].
\end{eqnarray}
The correlation function becomes
\begin{equation}\label{debath}
  \langle \hat B\left(t\right)\hat B\left(t-\tau\right)\rangle_B=\sum_{p,q\neq p}\kappa^2\hat n_p\left(1+\hat n_q\right)\exp\{i(\nu_p-\nu_q)\tau\}\,.
\end{equation}
As before, we now extend the $\tau$ integration to infinity and in Eq.\,(\ref{debath}) we change the summation over $p$ ($q$) with the integral, $\sum_{p(q)}\to\int_0^\infty d\nu_{p(q)}\,g_k(\nu_{p(q)})$, for example
\begin{eqnarray}
   &&\int_0^t  d\tau\, \langle \hat B^\dagger(t)\hat B(t-\tau)\rangle_B\to \nonumber \\ \nonumber
   &&\int_0^{\infty} d\nu_p d\nu_q\,g_k\left(\nu_p\right)g_k\left(\nu_q\right)\kappa^2\left(\nu\right)\bar{n}\left(\nu_p,T\right)\left[1+\bar{n}\left(\nu_q,T\right)\right] \\
   &\times & \int_0^\infty d\tau\,e^{i\left(\nu_p-\nu_q\right)\tau}\,.
\end{eqnarray}
The time integral is $\int_0^\infty d\tau\,e^{i\left(\nu_p-\nu_q\right)\tau}=\pi\delta(\nu_p-\nu_q)+i\mathcal{P}/(\nu_p-\nu_q)$. We omit here the contribution of the terms containing the Cauchy principal value $\mathcal{P}$, but they must be included in the Lamb-shifted Hamiltonian. Transforming back to the Schr\"{o}dinger picture, we obtain the pure dephasing contribution to the master equation for a generic system in thermal equilibrium
\begin{equation}
   \dot{\hat\rho}=\sum_k\gamma^{(k)}_{\rm \varphi}\mathcal D\left[ \sum_m s^{(k)}_{mm} \vert m\rangle\langle m\vert\right]\hat\rho
   \label{PureD}
\end{equation}
with 
\begin{equation}
  \gamma^{(k)}_{\varphi}=2\pi\int_0^\infty d\nu\,\kappa^2(\nu)g_k^2(\nu)\bar n(\nu,T)\left[1+\bar n(\nu,T)\right]\,.
\end{equation}
Using Eq.\,(\ref{Relax}) and (\ref{PureD}), we obtain the master equation valid for generic hybrid-quantum systems in the weak-, strong-, ultra-strong coupling regime, with or without parity symmetry.

\section{Dynamical Decoupling performance }\label{appDD}
In a pure dephasing picture, a two-level system is described by
\begin{equation}
  \hat H=\left(\frac{\omega_q}{2}+\beta\left(t\right)\right)\hat\sigma_z\,,
\end{equation}
where $\omega_q$ and $\beta(t)$ represent the energy transition and random fluctuations imposed by the environment.
The frequency distribution of the noise power for a noise source $\beta$ is characterized by its power spectral density
\begin{equation}
  S\left(\omega\right)=\frac{1}{2\pi}\int_{-\infty}^{\infty} dt\langle\beta\left( 0\right) \beta\left( t\right)\rangle e^{-i\omega t}
\end{equation}
The off-diagonal elements of the density matrix for a superposition state affected by decoherence is 
\begin{equation}
\rho_{01}(t)=\rho_{01}(0)\exp{[-i\Sigma(t)]}\exp{[-\chi(t)]}	\,.
\end{equation}
The last term is a decay function and generates decoherence, it is the ensemble average of the accumulated random phase $\exp{[-\chi(t)]}=\langle\exp{[i\delta\varphi(t)]} \rangle$, with $\delta\varphi(t)=\int_0^t d{t}'\delta\beta{({t}')}$. Following Ref.\,\cite{Uhrig:2007}, we have that
\begin{equation}
  \chi\left(\tau\right)=\int_0^\infty d\,\omega S\left(\omega \right)\frac{F\!\left(\omega t\right)}{\omega^2} \coth{\left(\frac{\hbar\omega}{2k_B T}\right)}\,.
  \label{chi}
\end{equation}
When the system is free to decay, free induction decay (FID), then $F(\omega t)=2 \sin{(\omega t/2)}^2$. If we apply a sequence of $N$ pulses, then $F(\omega t)=\vert Y_N(\omega t) \vert^2/2$, with
\begin{equation}
  Y_N(z)=1+\left(-1\right)^{N+1}\exp\{iz\}+2\sum_{j=1}^{N}\left(-1\right)^j\exp\{iz\delta_j\}\,.
  \label{YN}
\end{equation}
Using superconducting artificial atoms, the power spectral density exhibits a $1/f$ power-law, $S(2\pi f)=A/f$, where $A$ is a parameter that we will evaluate assuming to know the pure dephasing time of the system during FID. Indeed, we calculate the integral $\chi_0=\chi(\tau_{\rm FID})$ in Eq.\,\ref{chi}, considering that the pure dephasing time is $\tau_{\rm FID}=10\,\mu \rm s$ and $A=1$. After that we choose $A=1/\chi_0$, in $S(2\pi f)$. With this choice of $A$, we are sure that, $\exp{[-\chi(\tau_{\rm FID})]}=1/e$, and that the pure dephasing rate, when the system is free to decay, is $\Gamma_{\rm FID}=1/\tau_{\rm FID}$.
At this point, we can calculate $\chi_N=\chi(\tau)$ in Eq.\,\ref{chi} for a sequence of $N$ equidistant pulses, $\delta_j=j/(N+1)$, using Eq.\,\ref{YN} and $A=1/\chi_0$. If $\alpha_{N}$ is the pure dephasing suppression factor, $\Gamma_N=\alpha_N\Gamma_{\rm FID}$, it results that $\alpha_N=\sqrt{\chi_N}$.
Considering $\tau_{\rm FID}=10\,\mu \rm s$ and $T=12\,$mK, we found $A=4.34\times 10^{9}$. Applying $1000$ equally spaced pulses, the suppression factor is $\alpha_N= 10^{-3}$. In conclusion, applying a DD sequence of 1000 $\pi$-pulses in a two-level artificial atom that experiences noise with $1/f$ power spectral density, at low temperature the decoherence time can be prolonged up to $10^{3}$ times.

\section{Conditions for an auxiliary non-interacting atomic level}\label{auxiliary}
The frequency transitions between the auxiliary level $\vert s\rangle$ and the lowest two levels must be much greater than the one between the lowest two levels; this is facilitated by using a flux qubit in its optimal point. More importantly, the transition matrix elements between the auxiliary level and the lowest two levels should be much lower than the transition matrix element between the lowest two levels. For example, for a coupling $\lambda/\omega_c=1$, the transition matrix elements between the auxiliary level and the lowest two levels should be less than $10\%$ of the transition matrix element between the lowest two levels. In the case of longitudinal coupling, the matrix elements must be calculated between the states $\vert ge_\pm\rangle=(\vert g\rangle\pm\vert e\rangle)/\sqrt{2}$ and between the states $\vert es_\pm\rangle=(\vert e\rangle\pm\vert s\rangle)/\sqrt{2}$ and $\vert gs_\pm\rangle=(\vert g\rangle\pm\vert s\rangle)/\sqrt{2}$.
If, for some parameters, the last condition is not satisfied, another way to store the information would be to prepare the system in the state $\vert s\rangle$ when the coupling is low, $\lambda/\omega_c\leq 0.1$, and, after that the flying qubit enters the cavity, switching-on the coupling \cite{Peropadre:2010}. Afterwards, we follow the protocol described in the part of the main paper. To release the quantum information, we reverse the process.


\bibliography{QMinDSC}

\end{document}